# ClusterChirp: Scalable Interactive Exploration of Omics Data with Natural Language–Guided Analysis

Osho Rawal[1], Rex Lu[1], Edgar Gonzalez-Kozlova[2,3,4], Sacha Gnjatic[2,3], Zeynep H. Gümüş[1,3,4,*]

[1] Department of Genetics and Genomics, Icahn School of Medicine at Mount Sinai, New York, NY 10029, USA
[2] Department of Immunology & Immunotherapy, Icahn School of Medicine at Mount Sinai, New York, NY 10029, USA
[3] Precision Immunology Institute, Icahn School of Medicine at Mount Sinai, New York, NY 10029, USA
[4] Windreich Department of Artificial Intelligence and Human Health, Icahn School of Medicine at Mount Sinai, New York, NY 10029, USA

* To whom correspondence should be addressed. Tel: +1 212 824 8954; Fax: +1 212 241 3310; Email: zeynep.gumus@mssm.edu

## SUMMARY

High-dimensional omics datasets are routinely visualized as heatmaps, where color intensities reveal co-expression patterns and correlations. However, modern omics technologies increasingly generate matrices so large that existing visual exploration tools require down-sampling or filtering, causing loss of biologically important patterns. Additional barriers arise from tools that require command-line expertise, or fragmented workflows for downstream biological interpretation. We present ClusterChirp, a web-based platform for real-time exploration of large-scale data matrices. The platform combines GPU-accelerated rendering and parallelized hierarchical clustering using multiple CPU cores. Built on deck.gl and multi-threaded clustering algorithms, ClusterChirp supports on-the-fly clustering, multi-metric sorting, feature search and interactive visualization controls within a single interface. Uniquely, a natural language interface powered by a Large Language Model allows users to perform complex operations and build reproducible workflows through conversational commands. ClusterChirp further enables within-cluster correlation network analysis in 2D or 3D, and integrates functional enrichment through biological knowledge bases. Developed with iterative user feedback and adhering to FAIR4S principles, ClusterChirp enables users to extract insights from high-dimensional omics data with unprecedented ease and speed. It is freely available at clusterchirp.mssm.edu without login and is also distributed as a Dockerized application at ghcr.io/gumuslab/clusterchirp.

## INTRODUCTION

Modern omics technologies routinely generate high-dimensional datasets, profiling tens of thousands of molecular features across hundreds to thousands of samples (1). Transcriptomic, proteomic, and metabolomic experiments capture molecular landscapes under diverse biological conditions, generating rich repositories for systems-level discovery (2). However, the rapid growth in data generation has outpaced the capabilities of current computational infrastructure and visualization tools (3–5), creating a widening gap between data generation and biological interpretation.

A major bottleneck is data visualization. Clustering combined with heatmap visualizations is widely used to identify co-regulated molecular features and global patterns (6), but these methods scale poorly. Hierarchical clustering typically exhibits quadratic to cubic time complexity (7), requiring tens of millions of pairwise distance calculations for datasets with ~10,000 genes, making real-time, interactive visualization infeasible on standard hardware. Web-based tools introduce additional constraints, including session timeouts, server-side processing limits, and memory caps, often forcing users to down-sample data and potentially miss biologically meaningful patterns. Accessibility further compounds the issue, as many high-performance tools require command-line expertise or specialized syntax, limiting adoption among researchers without computational training.

Even when meaningful visual patterns are detected, downstream interpretation often relies on fragmented workflows that require exporting gene lists, converting identifiers, and manually querying external enrichment tools. These steps slow analysis, increase the potential for error, and can lead to incomplete biological interpretation. Numerous heatmap visualization tools have been developed, including web-based platforms (8–13) that offer varying degrees of interactivity and enrichment support. While widely used, these tools often face limitations in scalability, session stability, computational throughput, or long-term availability when applied to large-scale omics datasets. Desktop applications and programming libraries, including MATLAB, R and Python packages such as ComplexHeatmap (14), pheatmap (15), seaborn (16) , and plotly (17) provide greater flexibility, but remain constrained by memory limits, licensing restrictions, or usability barriers. As a result, large-scale omics studies are frequently limited to static visualizations, slow exploratory cycles, or partial functional interpretation.

These limitations impede biological discovery. Down-sampling can mask important functional subgroups, while static or slow visualizations hinder iterative hypothesis generation, such as adjusting clustering parameters, filtering by metadata, or updating visualizations in real time. Functional interpretation and within-cluster relationship analyses, including correlation networks, often require separate tools (18–20), further fragmenting the workflows and increasing cognitive overhead. Together, these challenges highlight an unmet need for scalable, interactive visualization platforms that integrate downstream biological interpretation while remaining accessible to users with varying levels of computational expertise.

Large language models (LLMs) offer a complementary approach to improving usability. By enabling natural language interaction, LLMs allow users to issue high-level commands, such as "*cluster genes using Pearson correlation*", without navigating complex interfaces or writing code (21,22). While LLMs do not address computational scalability, they reduce barriers for non-programmers and accelerate exploratory workflows for experienced users (23,24).

To address both computational and usability challenges, we developed **ClusterChirp**, a GPU-accelerated web-based platform for real-time interactive exploration of omics datasets containing thousands of samples and tens of thousands of biomarkers. The platform scales to datasets approaching 10 million cells while maintaining responsive performance. ClusterChirp integrates several key features. GPU-based rendering using WebGL (leveraging deck.gl for heatmaps and Sigma.js/Three.js for network visualizations) enables smooth interaction with large-scale data, while computational workloads are handled on the server. A parallelized hierarchical clustering engine distributes distance calculations across CPU cores, reducing runtimes from hours to minutes. The platform also features an LLM-powered natural language interface for conversational analysis, supports within-cluster pairwise correlation networks in both 2D and 3D, and provides real-time functional interpretation through integration with biological databases. By combining scalable computation, interactive visualization, functional interpretation, and accessible interfaces, ClusterChirp enables users to fully explore and interpret high-dimensional omics data. The tool is freely available as a public web server and can also be deployed locally via Docker.

## MATERIALS AND METHODS

**User-Centered Design and Prototyping.** ClusterChirp was developed using an iterative, user-centered design approach involving 10 domain practitioners in immunology, genomics and bioinformatics. Initial interviews with 5 experts identified three recurring issues limiting large-scale data exploration: i) the need to down-sample large datasets, risking loss of biologically important patterns; ii) processing delays exceeding 10 minutes for routine clustering tasks; and iii) limited support for metadata-aware filtering. Guided by these requirements, we engineered ClusterChirp to prioritize scalability, responsiveness, and usability. For user interface (UI) design, based on the expert feedback, we adopted the basic design elements from Clustergrammer (13), a commonly used heatmap tool. Feature development was guided by low-fidelity prototypes, followed by high-fidelity interactive implementations in React (25) and TypeScript (26). Biweekly semi-structured evaluation sessions with domain experts involved performing representative analysis tasks and providing real-time feedback, which informed iterative refinement of the UI and core functionalities, including the natural language interface and correlation network visualizations. The final prototype was evaluated by 5 additional researchers from Mount Sinai and Brown University to ensure usability and relevance to large-scale datasets.

**FAIR4S Principles and Software Availability.** To ensure sustainability, ClusterChirp was developed following Findable, Accessible, Interoperable, Reusable for Research Software (FAIR4S) principles.
Findable: Available at a stable URL, with version source code hosted on GitHub (frontend: https://github.com/GumusLab/HeatmapFrontend; backend: https://github.com/GumusLab/HeatmapBackend)
Accessible: Freely accessible via modern web browsers without registration or installation; open-source under APGL-3.0 license.
Interoperable: Supports input files in TSV, CSV, and Microsoft Excel formats; exports results as TSV and PDF.

Reusable: Documentation, tutorials, and example datasets with expected outputs are provided to support transparent reuse and reproducibility.

**Interactive Heatmap Visualization.** ClusterChirp enables real-time visualization of large heatmaps with integrated metadata and dendrograms using GPU-accelerated rendering via the WebGL-based framework deck.gl (27). The heatmap is built as a custom React component, with matrix cells rendered using ScatterLayer, and annotations rendered using TextLayer. To optimize performance, matrix data are stored as Float32 arrays for efficient GPU data transfer, and viewport-based culling renders only visible regions during visual navigation. Interactive features, including tooltips, pan, zoom, and search highlighting are managed through deck.gl event handlers, while computationally intensive tasks such as sorting and filtering are offloaded to Web Workers to maintain UI responsiveness. The GPU is used strictly for rendering, while clustering and correlation computations are performed on server CPUs (see Backend Architecture and Clustering Implementation).

**Natural Language Interface (AI Chatbot).** ClusterChirp includes a natural language chatbot powered by an LLM (GPT-4o-mini version gpt-4o-mini-2024-07-18; https://platform.openai.com/docs/models/gpt-4o-mini), enabling users to perform operations through conversational commands. The system interprets commands and translates them into structured actions applied to the active visualization state, including filtering, sorting, clustering, and enrichment queries. Commands are categorized as front-end operations (e.g. adjust opacity; sort); backend filtering and/or clustering (e.g. filter by metadata fields); and enrichment queries, which invoke curated gene sets before applying filters to the dataset. Frontend commands update visualization layers directly, while backend commands are processed via a Django API. For each command, a dynamic prompt is constructed, including the current query, session command history, active clustering state, filtering settings, and session-specific metadata (**`session_id_metadata.json`**). This enables context-aware processing, allowing sequential commands to build on prior operations. The LLM returns a standardized JSON action object, which is validated against dataset parameters prior to execution. Invalid values trigger error messages listing available alternatives, and a rule-based fallback parser ensures functionality when the LLM API is unavailable.

**Correlation Network Visualization.** For any user-selected cluster, ClusterChirp enables building within-cluster correlation networks to reveal co-expression patterns among its members. Networks can be rendered in either 2D or 3D. Prior to network construction, biomarkers with >20% missingness are excluded (28), and analysis is restricted to high-variance biomarkers (default: top 75%) (29). Pairwise similarity is then computed using Pearson correlation. To scale to large biomarker sets, computations are parallelized using the *joblib* library (https://joblib.readthedocs.io/) and accelerated with Numba JIT compilation (https://numba.pydata.org/). Statistical significance of each correlation is assessed using Fisher's z-transformation for sufficiently large sample sizes (n nodes > 30); and a t-distribution approximation otherwise. By default, only strong, significant associations are retained ($|r| \geq 0.7$, and $p<0.05$) (30), though these thresholds are user-adjustable. The resulting network is truncated if over N=100,000 edges, with each edge annotated by correlation coefficient, p-value, and sample size, and stored in JSON format.

For visualizations, 2D networks are rendered using Sigma.js (v3.0) (https://www.sigmajs.org/) with WebGL acceleration and laid out via the force-directed ForceAtlas2 (31) algorithm, which spatializes nodes based on correlation strength. Nodes are colored by their connectivity degree to highlight hubs. In 3D, networks are rendered with Three.js (WebGL) (https://threejs.org/). Subclusters are first identified using Leiden community-detection algorithm (32), with each subcluster organized into distinct regions on a 3D sphere, with nodes colored by subcluster membership. For large networks (>1,000 nodes), layout and clustering computations are run in browser-based Web Workers to maintain UI responsiveness. Both 2D and 3D viewers support real-time interactivity, including node search, autocomplete, highlighting, with efficient GPU-based re-rendering that preserves a fluid user experience.

**Gene-set enrichment analysis.** Upon selection of a cluster, gene identifiers are extracted and harmonized to standardized HUGO gene symbols. For Olink Target proteomics panels (33), protein names are mapped to gene symbols using a predefined dictionary (https://olink.co). ClusterChirp supports two complementary workflows for gene-set enrichment analysis, both leveraging Enrichr (34–36). In the first workflow, the standardized gene list is submitted to Enrichr, and results are opened in a new browser tab via the native Enrichr interface. In the second workflow, enrichment analysis is performed directly within ClusterChirp, with results rendered in an integrated, interactive view without leaving the platform.

**Tutorials.** A four-module tutorial series covers basic navigation and heatmap interaction; natural language interaction; enrichment analysis, and correlation network exploration in 2D and 3D. Each module includes step-by-step instructions, annotated screenshots, demo videos, example datasets, and troubleshooting tips for common issues. The tutorials are fully responsive for use across desktops, tablets, and mobile devices.

**Data Security and Session Management.** ClusterChirp uses secure, session-based data handling implemented in Django on Mount Sinai's HIPAA-compliant Minerva server environment (37). Uploaded datasets are assigned a unique session ID and stored temporarily, ensuring isolation between users and sessions. Metadata (e.g. categorical annotations) are extracted and stored in lightweight JSON files to enable fast queries without loading full matrices. Django's session framework maintains a secure, persistent state, and temporary files are automatically removed on session expiration, ensuring data privacy and storage efficiency.

**Backend Architecture and Clustering Implementation.** ClusterChirp backend is built on Django 4.2 (https://www.djangoproject.com/) with Django REST Framework 3.15.2, utilizing pandas 2.0.3 (https://pandas.pydata.org/) and NumPy 1.24.4 (https://numpy.org/) for core data handling. Hierarchical clustering is performed via clustergrammer-py (13) integrated into custom REST API endpoints. To scale to large datasets (>20,000 biomarkers), pairwise distance computations are parallelized using *joblib* (https://joblib.readthedocs.io/) with adaptive caching (default 500 MB) and block-based computation to prevent memory overflow. This approach distributes the sequential $O(n^2)$ computation across available hardware resources (see Results for performance benchmarks). Users can select different distance metrics (Euclidean, Manhattan, Pearson), linkage methods (complete, average, single) via the API. Users can also select hierarchical groupings at multiple cutoff thresholds (0.0 to 1.0 in increments of 0.1) to explore dendrogram structures at different levels of granularity.

**Data Preprocessing and Imputation.** Missing values in uploaded datasets are handled through multiple imputation strategies implemented on the backend. Supported methods include mean and median imputation, k-nearest neighbors (KNN) imputation (38), iterative imputation (39), matrix factorization, correlation-based approaches, and random forest–based (40) imputation. These methods are implemented using established Python libraries, primarily scikit-learn (41) (KNNImputer, IterativeImputer, and ensemble-based models), with NumPy and pandas used for statistical operations. Matrix factorization and correlation-based approaches are implemented using custom routines built on NumPy. All imputation procedures are executed via Django REST API endpoints, with computations parallelized where applicable using joblib to ensure scalability for large datasets.

## RESULTS

**Platform overview**. ClusterChirp is a web-based platform for real-time exploration of large-scale omics data, integrating scalable heatmap visualization, natural language-driven analysis, and downstream functional interpretation within a single interface. It was developed using an iterative, user-centered development process involving domain experts, ensuring alignment with real-world omics analysis pipelines. By combining GPU-accelerated rendering with parallelized backend computation, the platform allows interactive analysis of datasets at scales that typically require offline processing or down-sampling. Core functionalities include dynamic heatmap visualization, natural language query execution, correlation network analysis, and gene set enrichment.

**Data ingestion and preprocessing.** Upon data upload ClusterChirp automatically validates input structure and detects missing values. If no missing values are present, clustering proceeds immediately and a full-screen heatmap with integrated metadata annotations and dendrograms is rendered. If missing values are detected, users are presented with summary statistics and are offered multiple imputation strategies, including mean or median replacement, k-nearest neighbors (KNN) (38), iterative imputation (39), matrix factorization, correlation-based methods, and random forest (40). Alternatively, an auto-select option recommends an imputation method based on dataset characteristics (e.g. missing data percentage, distribution patterns), enabling rapid preprocessing.

**Interactive visualization and navigation.** ClusterChirp provides a single interface for visualization, analysis, and interpretation (Figure 2A). Users can filter data, perform clustering, analyze within-cluster relationships as networks in 2D or 3D, and run enrichment analyses directly on the active dataset, with updates reflected across all views in real time. A control panel supports direct manipulation of the dataset, including sorting rows and columns (alphabetical, sum-based, or variance-based); hierarchical clustering, biomarker (row label) search and highlighting, opacity adjustment to emphasize extreme values and to adjust contrast, and metadata selection for display. Metadata-driven reordering can be triggered directly; for example, selecting a category such as “Sex” reorganizes the samples into groups in real time, revealing associated patterns without re-computation (Figure 2B). A search box with autocomplete enables rapid identification of specific biomarkers, automatically centering and highlighting selected rows. To support navigation of large datasets, a minimap provides a global overview, allowing users to reposition the main view by directly clicking on and dragging a viewpoint indicator, enabling rapid repositioning of the main view while preserving spatial context. Users can export high-resolution PDF snapshots suitable for publication (with the option to turn on/off the minimap), download processed datasets as TSV files, and crop regions of interest for focused explorations.

**Real-time hierarchical clustering.** ClusterChirp supports dynamic reorganization of rows and columns through hierarchical clustering using multiple distance metric options (cosine, Euclidean, Manhattan, Pearson) and linkage methods (complete, average, single). Clustering is performed through a parallelized backend pipeline, enabling datasets with tens of thousands of rows to be processed within minutes.

**Metadata integration.** ClusterChirp automatically extracts and metadata as annotation layers (Figure 2A). Column metadata are displayed as horizontal bars above the heatmap, and row metadata as vertical bars to the left. Up to three metadata tracks per axis can be displayed simultaneously and updated dynamically through the control panel. Following clustering, dendrograms are rendered alongside the matrix (Figure 2B), and interactive sliders allow users to adjust clustering depth in real time (for both rows and columns), facilitating exploration of hierarchical structure at multiple resolutions (Figure 2B). Hovering over dendrogram branches provide contextual summaries of metadata distributions within clusters.

**Integration with functional enrichment tools.** Functional interpretation is supported through direct integration with the Enrichr platform (34–36). Users can submit to Enrichr selected gene or protein lists with a single click, and protein identifiers are automatically mapped to HUGO gene symbols prior to analysis. This enables rapid interpretation of clustered rows in the context of known biological pathways. Additionally, ClusterChirp integrates multiple pathway databases (KEGG, Reactome, WikiPathways, MSigDB), supporting pathway discovery, functional filtering, and targeted pathway selection through the natural language chatbot or in select subclusters within the correlation networks as described below.

**Natural language-driven data exploration.** ClusterChirp includes a conversational chatbot powered by an LLM that enables users to perform analysis operations using natural language commands. Supported operations include filtering, clustering, sorting, feature selection, search, and visualization adjustments. Commands range from simple queries (e.g. “show only male samples”) to multi-step operations (e.g. "cluster genes using Pearson correlation and average linkage") (Supplementary Table S1). Users can filter rows or columns based on metadata (e.g., "show only female samples"), combine multiple conditions (e.g., "show female who are alive"), with all applied filters displayed as removable tags for transparency. Command suggestions are organized by functional category, and command history is preserved across a session to support iterative exploration. Together, these features offer a practical alternative to traditional GUI interactions while lowering the barrier to complex analyses.

**Natural language-driven pathway enrichment**. Using the LLM chatbot, users can filter the heatmap for genes or proteins based on biological pathways and functional categories. Commands such as “list immune pathways” or “show cancer pathways” will reveal what categories are available for query; broad queries such as “show immune genes” or “metabolism genes” will automatically subset and re-cluster the dataset; and targeted queries for specific pathways or identifiers such as “filter by pathway KEGG_MAPK_signaling”, will generate focused heatmaps that reveal pathway-specific patterns that may not be apparent in the full dataset.

**Natural language interface evaluation.** We benchmarked the natural language interface using 45 representative commands spanning filtering (n=20), clustering (n=14), and sorting (n=11) (Supplementary Table S1). Success was defined as correct execution without manual intervention, while commands requiring rephrasing were recorded separately. Response time was measured from command submission to completion of the requested operation. Testing on a bladder cancer immunotherapy dataset (42) (Case Study 2 below) demonstrated high execution accuracy across supported operations (95.66%), with average response times under 2 seconds, low fallback rates, and minimal need for command rephrasing (Supplementary Table S2).

**Correlation network analysis, pathway enrichment and visualization.** To facilitate the interpretation of clustered rows, ClusterChirp enables correlation-based network analysis within user-selected clusters. Pairwise similarity is computed among cluster members to identify statistically significant associations, which are visualized as interactive networks in 2D or 3D layouts. These network representations help identify densely connected modules that may reflect co-expression patterns, shared biological pathways, or regulatory relationships, complementing hierarchical clustering with an additional structural perspective. Selecting a cluster within the dendrogram opens a dialog displaying cluster statistics and analysis options; selecting network generation launches an interactive 2D network view in which nodes represent biomarkers (rows) and edges denote statistically significant pairwise correlations among cluster members. Nodes are color-coded by connectivity (dark blue >10 connections; medium blue: 5-10 connections; light blue: <5 conditions). The interface includes controls for row ID search and label display. Leiden clustering can be applied to identify community structure, which can be visualized in a 3D network view. Users can perform gene set enrichment analysis on selected clusters using the “Analyze Pathways” option. This launches an interactive enrichment module within ClusterChirp, where significantly enriched pathways are displayed. Users also have the option to open the full analysis in Enrichr (34–36) via a dedicated link, which directs the results to a new browser tab.

**CASE STUDIES.** We demonstrate ClusterChirp’s capabilities using two proteomics studies, which are both available for interactive exploration on the web platform.

**Case Study 1: Spatial proteomics data analysis with ClusterChirp reveals distinct cell-type specific signatures in the tumor microenvironment.** We analyzed multiplexed immunohistochemistry (mIHC) data from six cancer patients (43), comprising single-cell protein measurements with cells annotated into eight major lineages based on marker expression (cancer cells, CD8+ T cells, CD8- FOXP3- T cells, regulatory T cells, B cells, plasma cells, macrophages, stromal cells). The dataset included eight protein markers (CD20, CD3, CD68, CD8, FOXP3, MZB1, PanCK, aSMA), each quantified at four intensity levels (p1, p5, p9, mean), yielding 32 features per cell. We first generated a representative subset of 1,202 cells (~200 per patient) using proportional sampling with a separation score that prioritizes prototypical cells. Next, ClusterChirp hierarchical clustering separated the eight cell types into distinct groups (Figure 3A). Related immune cell subtypes clustered together (CD8+ T cells, FOXP3+ T cells, CD8+ FOXP3+ T cells, and CD3+ T cells), while cancer cells (PanCK+), B cells (CD20+/MZB1+), macrophages (CD68+), and stromal cells (aSMA+) formed distinct branches, with clear differential expression patterns (red: high; blue: low). Clicking on a cluster layer branch opened an interactive dialog with options to explore within-cluster pathway enrichment results, visualize the within-cluster correlation network, or to directly upload the selected cluster’s list of biomarkers to Enrichr (34–

36). Choosing the network visualization revealed cell type community structure, where the network view (Figure 3B) displayed all 1,202 cells as nodes colored by cell type, with spatial positioning based on marker expression similarity. This visualization clearly shows the separation of distinct cell populations into communities, with T cell subtypes clustering together while other cell types forming separate groups.

Next, we used sub-setting and re-clustering options to verify that cluster assignments reflect true biological differences rather than technical artifacts. Towards this end, to identify both major cell lineages and subtle subpopulations within the tumor microenvironment, we explored the heatmap at multiple scales using the dendrogram depth slider (Figure 3C). Traversing across 11 hierarchical levels revealed progressively finer cell type distinctions at deeper cuts. Then, we explored specific populations by selecting and cropping B cells for detailed examination (Figure 3D). The cropped view showed high CD20 expression across all cells in this cluster, confirming B cell identity. In this case study, having the options to adjust the dendrogram depth, and analyze subsets of specific cell populations through co-expression network views made it easy to visually confirm that the observed separations reflected actual biological differences in protein expression rather than technical artifacts or batch effects. This interactive validation approach is especially useful for spatial proteomics data, where visual inspection of marker expression patterns is essential for confirming cell type assignments and identifying potential misclassifications.

**Case Study 2. ClusterChirp enables natural language-assisted analysis of bladder cancer treatment response biomarkers.** We next analyzed longitudinal plasma proteomics data from a bladder cancer immunotherapy trial (GU16-257; data kindly provided by the study investigators) (42). The dataset includes 77 proteins from the Olink Immuno-Oncology panel (after QC filtering), measured across treatment cycles (C1D1, C3D1, C8D1, C12D1; C = cycle, D = day) in 196 samples, and reported as log2-transformed Normalized Protein eXpression (NPX) values. Associated metadata includes patient ID (~50 unique patients), timepoint, response (yes, no, NE), gender (M, F), race (White, Asian, Black, Unknown), ethnicity (Non-Hispanic, Hispanic, Unknown).

Hierarchical clustering identified distinct co-expression clusters (Figure 4A). The dendrogram on the left margin showed clear separation between clusters, with Cluster 2 containing 42 genes displaying coordinated expression patterns. We customized the visualization by selecting metadata bars to display the heatmap, showing response status (responder/non-responder), timepoint, and patient ID for each sample, with red indicating higher protein expression and blue indicating lower expression. Using the natural language interface, we filtered samples by treatment timepoint ("Select samples at C3D1 timepoint"). The system filtered the heatmap in real time and displayed data only from C3D1 samples (Figure 4B). Double-clicking on the *FASLG* gene sorted all samples by this gene's expression values in descending order, and revealed that patients who achieved complete clinical response (dark blue response metadata track) were enriched on the left side of the heatmap where *FASLG* expression was highest (Figure 4B), aligning with published findings from this trial linking high *FASLG* expression to better treatment outcomes.

To interpret the biological functions of proteins co-expressed with *FASLG*, we selected Cluster 2 (Figure 4B) and clicked on "ANALYZE PATHWAY". This submitted the cluster genes list to the external Enrichr tool (34–36), for enrichment analysis against its curated library collection, with results displayed within ClusterChirp. From the KEGG 2021 Human pathway database (Figure 4C), the top enriched pathway was cytokine-cytokine receptor interaction (p-value = 2.25e-10), indicating that proteins co-expressed with *FASLG* are predominantly involved in immune cell communication and signaling. Other enriched pathways included *viral protein interaction with cytokines*, *TNF signaling*, and *chemokine signaling pathways*, collectively suggesting coordinated immune and inflammatory responses. This workflow, from filtering and clustering to interpretation was performed without writing any code, demonstrating how ClusterChirp supports rapid, interactive exploration and hypothesis generation in complex datasets.

**Scalability, performance and benchmarking.** ClusterChirp is deployed on Mount Sinai's HIPAA-compliant Minerva infrastructure (37), with session-based data handling to ensure data isolation. As clustering requires computation of an $n \times n$ distance matrix ($O(n^2)$ memory), memory requirements roughly scale with the square of the number of rows. Practical dataset size therefore depends on available system memory: systems with 16 GB RAM typically support datasets up to approximately 30,000 rows, with ~20,000 rows recommended as a conservative guideline for routine use.

Benchmarking against five widely used web-based tools demonstrated that ClusterChirp supports larger datasets while maintaining interactivity (see Table 1). These tools include Morpheus, NG-CHM, Heatmapper2 (10), ClustVis (11), and Clustergrammer (13). Using a synthetic dataset designed to test scalability limits (5,000 × 2,000; 10 million cells) and a real-world proteomics dataset (10,909 proteins × 108 samples; ~1.2 million cells), ClusterChirp consistently completed clustering within minutes while preserving interactive visualization (Table 1). In contrast, several tools failed due to dataset size limits or server constraints. Briefly, for the synthetic dataset, ClusterChirp completed clustering in ~3 minutes, Heatmapper2 in ~30 seconds, and NG-CHM in 7+ minutes. Morpheus remained incomplete after 12+ minutes, ClustVis rejected the file due to size limits, and Clustergrammer returned a 502 server error. For the real-world dataset, ClusterChirp completed clustering in 30 seconds and Heatmapper2 in ~25 seconds, while NG-CHM and ClustVis rejected the dataset due to row limits; Morpheus failed; and Clustergrammer returned server errors. Although Heatmapper2 achieved fast runtimes by running computations client-side via WebAssembly and avoiding network latency, it generates static images that do not support interactive exploration. In contrast, ClusterChirp maintains responsive visualization while scaling to large datasets. This performance advantage is driven by parallelized pairwise distance computation, which yields over a 3-fold speedup for datasets with >20,000 biomarkers compared to sequential approaches, without loss of accuracy (see Methods). As a result, users can analyze large datasets without down-sampling, preserving biologically relevant patterns.

In addition, we evaluated ClusterChirp performance with datasets ranging from small gene panels (100 genes × 50 samples) to large-scale omics datasets (23,563 genes × 478 samples; ~11 million cells). Across datasets ranging up to ~10 million cells, visualization remained interactive (~60fps pan/zoom), with gradual performance decline beyond this scale due to browser memory limits. Clustering completed in minutes rather than hours across all tested dataset sizes. These results demonstrate that ClusterChirp enables interactive exploration of large omics datasets without requiring down-sampling. Cross-browser compatibility was confirmed on Chrome, Safari, and Firefox.

## Discussion

We present ClusterChirp, an interactive web platform for exploratory analysis of high-dimensional omics data that integrates visualization, clustering, and downstream interpretation within a single environment. By combining GPU-accelerated rendering with parallelized computation and natural language-guided interaction, it enables real-time, iterative analyses that are often fragmented across multiple tools or require offline preprocessing. A natural language chatbot enables users to run analyses via simple commands, such as clustering genes by expression levels, reducing the need to navigate complex operations or command-line instructions. This lowers the barriers to entry for non-programmers while preserving flexibility for more advanced users.

ClusterChirp further facilitates biological interpretation via its built-in pathway enrichment through Enrichr (34–36), allowing researchers to dive directly from clustering results to functional insights without exporting data or switching platforms. Rather than redirecting users to an external website, ClusterChirp calls the Enrichr API directly and displays results within the interface, maintaining a continuous and streamlined workflow. Overall, by blending user-friendly interfaces with top-tier computing power, ClusterChirp sets a new standard for tools that let every researcher, no matter their technical skills, perform interactive, real-time studies on massive omics datasets. To maximize usability while ensuring data security, ClusterChirp is freely available at https://clusterchirp.mssm.edu, and deployed on Mount Sinai's HIPAA-compliant Minerva infrastructure and is also available as a Dockerized application at ghcr.io/gumuslab/clusterchirp.

This tool should be considered in the context of its limitations. First, the current hierarchical clustering implementation has computational constraints that limit analysis of very large datasets such as single-cell sequencing data containing hundreds of thousands of rows. Future development will focus on integrating computationally simpler clustering algorithms to address this limitation. Second, natural language interpretation, while robust in our evaluation, may require rephrasing ambiguous or complex queries. Third, pathway enrichment functionality depends on the Enrichr API (34–36); if this service is unavailable, enrichment analysis will be temporarily affected. Fourth, performance depends on Sinai's Minerva computing infrastructure; server downtime or high load may impact response times. Finally, as with any tool that accepts user-uploaded data, incorrectly formatted uploads may cause errors, though the platform provides format guidelines to minimize this issue.

### Data and code Availability

ClusterChirp source code is available on GitHub, with front end at https://github.com/GumusLab/HeatmapFrontend and heatmap backend at https://github.com/GumusLab/HeatmapBackend. Both repositories are released under the GNU Affero General Public License version 3.0 (GNU AGPL-3.0, https://www.gnu.org/licenses/agpl-3.0.en.html) and are also available under a commercial license for enterprises seeking additional features or wishing to avoid AGPL obligations. The tool is freely available at clusterchirp.mssm.edu and is also distributed as a Dockerized application at ghcr.io/gumuslab/clusterchirp.

### Author contributions

O.R., writing – original draft, writing – review & editing, visualization, software, and investigation; R.L., visualization, and software; S.G., writing – review & editing, conceptualization, and visualization;E.G.K., writing – review & editing, conceptualization, and visualization; Z.H.G., writing – original draft, writing – review & editing, conceptualization, visualization, investigation, and supervision.

### Acknowledgments

Authors gratefully acknowledge funding support from NCI R33 CA263705 to ZHG. This work was supported in part through the Minerva computational and data resources and staff expertise provided by Scientific Computing and Data at the Icahn School of Medicine at Mount Sinai and supported by the Clinical and Translational Science Awards (CTSA) grant UL1TR004419 from the National Center for Advancing Translational Sciences. Research reported in this publication was also supported by the Office of Research Infrastructure of the National Institutes of Health under award number S10OD026880 and S10OD030463. The content is solely the responsibility of the authors and does not necessarily represent the official views of the National Institutes of Health. We also thank Dr. Nicolas Fernandez, Dr. Alper Uzun, Dr. Dilber Ece Uzun, Dr.Seunghee Kim-Schulze, Darwin D'Souza, Dr. Myvizhi Esai Selvan, and Nitin Sreekumar for helpful discussions and feedback during development.

### Declaration of Interest

S.G. reports other research funding from Boehringer-Ingelheim, Bristol-Myers Squibb, Celgene, Genentech, Regeneron, and Takeda and consulting for Taiho Pharmaceuticals not related to this study.

### REFERENCES

**1.** Aebersold,R. and Mann,M. (2003) Mass spectrometry-based proteomics. Nature, 422, 198–207. https://doi.org/10.1038/nature01511.
**2.** Mahieu,N.G. and Patti,G.J. (2017) Systems-level annotation of a metabolomics data set reduces 25 000 features to fewer than 1000 unique metabolites. Anal. Chem., 89, 10397–10406.

https://doi.org/10.1021/acs.analchem.7b02380.
**3.** Mohr,A.E., Ortega-Santos,C.P., Whisner,C.M., Klein-Seetharaman,J. and Jasbi,P. (2024) Navigating challenges and opportunities in multi-omics integration for personalized healthcare. Biomedicines, 12, 1496. https://doi.org/10.3390/biomedicines12071496.
**4.** Subramanian,I., Verma,S., Kumar,S., Jere,A. and Anamika,K. (2020) Multi-omics data integration, interpretation, and its application. Bioinform. Biol. Insights, 14, 1177932219899051. https://doi.org/10.1177/1177932219899051.
**5.** Hasin,Y., Seldin,M. and Lusis,A. (2017) Multi-omics approaches to disease. Genome Biol., 18, 83. https://doi.org/10.1186/s13059-017-1215-1.
**6.** Eisen,M.B., Spellman,P.T., Brown,P.O. and Botstein,D. (1998) Cluster analysis and display of genome-wide expression patterns. Proc. Natl. Acad. Sci. U.S.A., 95, 14863–14868. https://doi.org/10.1073/pnas.95.25.14863.
**7.** Müllner,D. (2011) Modern hierarchical, agglomerative clustering algorithms. arXiv, arXiv:1109.2378.
**8.** Gould,J. (2016) Morpheus: versatile matrix visualization and analysis software. Broad Institute, Cambridge, MA, USA. https://software.broadinstitute.org/morpheus/.
**9.** Ryan,M.C., Stucky,M., Wakefield,C., Melott,J.M., Akbani,R., Weinstein,J.N. and Broom,B.M. (2019) Interactive clustered heat map builder: an easy web-based tool for creating sophisticated clustered heat maps. F1000Research, 8, 1750. https://doi.org/10.12688/f1000research.20590.1.
**10.** Babicki,S., Arndt,D., Marcu,A., Liang,Y., Grant,J.R., Maciejewski,A. and Wishart,D.S. (2016) Heatmapper: web-enabled heat mapping for all. Nucleic Acids Res., 44, W147–W153. https://doi.org/10.1093/nar/gkw419.
**11.** Metsalu,T. and Vilo,J. (2015) ClustVis: a web tool for visualizing clustering of multivariate data using principal component analysis and heatmap. Nucleic Acids Res., 43, W566–W570. https://doi.org/10.1093/nar/gkv468.
**12.** Ning,W., Wei,Y., Gao,L., Han,C., Gou,Y., Fu,S., Liu,D., Zhang,C., Huang,X., Wu,S., et al. (2022) HemI 2.0: an online service for heatmap illustration. Nucleic Acids Res., 50, W405–W411. https://doi.org/10.1093/nar/gkac480.
**13.** Fernandez,N.F., Gundersen,G.W., Rahman,A., Grimes,M.L., Rikova,K., Hornbeck,P. and Ma'ayan,A. (2017) Clustergrammer, a web-based heatmap visualization and analysis tool for high-dimensional biological data. Sci. Data, 4, 170151. https://doi.org/10.1038/sdata.2017.151.
**14.** Gu,Z., Eils,R. and Schlesner,M. (2016) Complex heatmaps reveal patterns and correlations in multidimensional genomic data. Bioinformatics, 32, 2847–2849. https://doi.org/10.1093/bioinformatics/btw313.
**15.** Kolde,R. (2010) pheatmap: Pretty heatmaps. R package version 1.0.12. https://doi.org/10.32614/CRAN.package.pheatmap.
**16.** Waskom,M.L. (2021) seaborn: statistical data visualization. J. Open Source Softw., 6, 3021. https://doi.org/10.21105/joss.03021.
**17.** Plotly Technologies Inc. (2015) Collaborative data science. Plotly Technologies Inc., Montréal, QC, Canada. https://plotly.com/.
**18.** Rawal,O., Turhan,B., Peradejordi,I.F., Chandrasekar,S., Kalayci,S., Gnjatic,S., Johnson,J., Bouhaddou,M. and Gümüş,Z.H. (2025) PhosNetVis: a web-based tool for fast kinase-substrate enrichment analysis and interactive 2D/3D network visualizations of phosphoproteomics data. Patterns, 6, 101148. https://doi.org/10.1016/j.patter.2024.101148.
**19.** Kalayci,S., Petralia,F., Wang,P. and Gümüş,Z.H. (2020) ProNetView-ccRCC: a web-based portal to interactively explore clear cell renal cell carcinoma proteogenomics networks. Proteomics, 20, e2000043. https://doi.org/10.1002/pmic.202000043.
**20.** Liluashvili,V., Kalayci,S., Fluder,E., Wilson,M., Gabow,A. and Gümüş,Z.H. (2017) iCAVE: an open source tool for visualizing biomolecular networks in 3D, stereoscopic 3D and immersive 3D. Gigascience, 6, 1–13. https://doi.org/10.1093/gigascience/gix054.
**21.** Wang,Q., Liu,X., Liang,M.Q., L'Yi,S. and Gehlenborg,N. (2023) Enabling multimodal user interactions for genomics visualization creation. In IEEE Visualization and Visual Analytics (VIS). IEEE, pp. 111–115. https://doi.org/10.1109/VIS54172.2023.00031.
**22.** Lange,D., Gao,S., Sui,P., Money,A., Misner,P., Zitnik,M. and Gehlenborg,N. (2023) YAC: bridging natural language and interactive visual exploration with generative AI for biomedical data discovery. [Preprint].

**23.** Shen,L., Shen,E., Luo,Y., Yang,X., Hu,X., Zhang,X., Tai,Z. and Wang,J. (2023) Towards natural language interfaces for data visualization: a survey. IEEE Trans. Vis. Comput. Graph., 29, 3121–3144. https://doi.org/10.1109/TVCG.2022.3148007.
**24.** Dibia,V. (2023) LIDA: a tool for automatic generation of grammar-agnostic visualizations and infographics using large language models. In Proc. 61st Annu. Meet. Assoc. Comput. Linguist., pp. 113–126. https://doi.org/10.18653/v1/2023.acl-demo.11.
**25.** React (2024) A JavaScript library for building user interfaces. Meta Platforms, Inc. https://react.dev/.
**26.** Microsoft Corporation (2024) TypeScript: JavaScript with syntax for types. Microsoft Corporation. https://www.typescriptlang.org/.
**27.** Uber Technologies, Inc. (2024) deck.gl: WebGL-powered framework for visual exploratory data analysis. Uber Technologies, Inc. https://deck.gl/.
**28.** Lazar,C., Gatto,L., Ferro,M., Bruley,C. and Burger,T. (2016) Accounting for the multiple natures of missing values in label-free quantitative proteomics data sets to compare imputation strategies. J. Proteome Res., 15, 1116–1125. https://doi.org/10.1021/acs.jproteome.5b00981.
**29.** Bourgon,R., Gentleman,R. and Huber,W. (2010) Independent filtering increases detection power for high-throughput experiments. Proc. Natl. Acad. Sci. U.S.A., 107, 9546–9551. https://doi.org/10.1073/pnas.0914005107.
**30.** Mukaka,M.M. (2012) Statistics corner: a guide to appropriate use of correlation coefficient in medical research. Malawi Med. J., 24, 69–71.
**31.** Jacomy,M., Venturini,T., Heymann,S. and Bastian,M. (2014) ForceAtlas2, a continuous graph layout algorithm for handy network visualization designed for the Gephi software. PLoS One, 9, e98679. https://doi.org/10.1371/journal.pone.0098679.
**32.** Traag,V.A., Waltman,L. and van Eck,N.J. (2019) From Louvain to Leiden: guaranteeing well-connected communities. Sci. Rep., 9, 5233. https://doi.org/10.1038/s41598-019-41695-z.
**33.** Chen,E.Y., Tan,C.M., Kou,Y., Duan,Q., Wang,Z., Meirelles,G.V., Clark,N.R. and Ma'ayan,A. (2013) Enrichr: interactive and collaborative HTML5 gene list enrichment analysis tool. BMC Bioinformatics, 14, 128. https://doi.org/10.1186/1471-2105-14-128.
**34.** Kuleshov,M.V., Jones,M.R., Rouillard,A.D., Fernandez,N.F., Duan,Q., Wang,Z., Koplev,S., Jenkins,S.L., Jagodnik,K.M., Lachmann,A., et al. (2016) Enrichr: a comprehensive gene set enrichment analysis web server 2016 update. Nucleic Acids Res., 44, W90–W97. https://doi.org/10.1093/nar/gkw377.
**35.** Xie,Z., Bailey,A., Kuleshov,M.V., Clarke,D.J.B., Evangelista,J.E., Jenkins,S.L., Lachmann,A., Wojciechowicz,M.L., Kropiwnicki,E., Jagodnik,K.M., et al. (2021) Gene set knowledge discovery with Enrichr. Curr. Protoc., 1, e90. https://doi.org/10.1002/cpz1.90.
**36.** Assarsson,E., Lundberg,M., Holmquist,G., Björkesten,J., Thorsen,S.B., Ekman,D., Eriksson,A., Rennel Dickens,E., Ohlsson,S., Edfeldt,G., et al. (2014) Homogenous 96-plex PEA immunoassay exhibiting high sensitivity, specificity, and excellent scalability. PLoS One, 9, e95192. https://doi.org/10.1371/journal.pone.0095192.
**37.** Kovatch,P., Gai,L., Cho,H.M., Fluder,E. and Jiang,D. (2020) Optimizing high-performance computing systems for biomedical workloads. IEEE Int. Symp. Parallel Distrib. Process. Workshops PhD Forum, 2020, 183–192. https://doi.org/10.1109/IPDPSW50202.2020.00040.
**38.** Troyanskaya,O., Cantor,M., Sherlock,G., Brown,P., Hastie,T., Tibshirani,R., Botstein,D. and Altman,R.B. (2001) Missing value estimation methods for DNA microarrays. Bioinformatics, 17, 520–525. https://doi.org/10.1093/bioinformatics/17.6.520.
**39.** van Buuren,S. and Groothuis-Oudshoorn,K. (2011) mice: multivariate imputation by chained equations in R. J. Stat. Softw., 45, 1–67. https://doi.org/10.18637/jss.v045.i03.
**40.** Stekhoven,D.J. and Bühlmann,P. (2012) MissForest—non-parametric missing value imputation for mixed-type data. Bioinformatics, 28, 112–118. https://doi.org/10.1093/bioinformatics/btr597.
**41.** Pedregosa,F., Varoquaux,G., Gramfort,A., Michel,V., Thirion,B., Grisel,O., Blondel,M., Prettenhofer,P., Weiss,R., Dubourg,V., et al. (2011) Scikit-learn: machine learning in Python. J. Mach. Learn. Res., 12, 2825–2830.
**42.** Galsky,M.D., Daneshmand,S., Izadmehr,S., Gonzalez-Kozlova,E., Chan,K.G., Lewis,S., El Achkar,B., Dorff,T.B., Cetnar,J.P., O'Neil,B., et al. (2023) Gemcitabine and cisplatin plus nivolumab as organ-sparing treatment for muscle-invasive bladder cancer: a phase 2 trial. Nat. Med., 29, 2825–2834. https://doi.org/10.1038/s41591-023-02568-1.
**43.** Buckup,M., Figueiredo,I., Ioannou,G., Ozbey,S., Cabal,R., Tabachnikova,A., Troncoso,L., Le Berichel,J., Zhao,Z., Ward,S.C., et al. (2025) Multiparametric cellular and spatial organization in cancer tissue lesions with a streamlined pipeline. Nat. Biomed. Eng. https://doi.org/10.1038/s41551-025-01475-9.

## LIST OF FIGURES AND TABLES

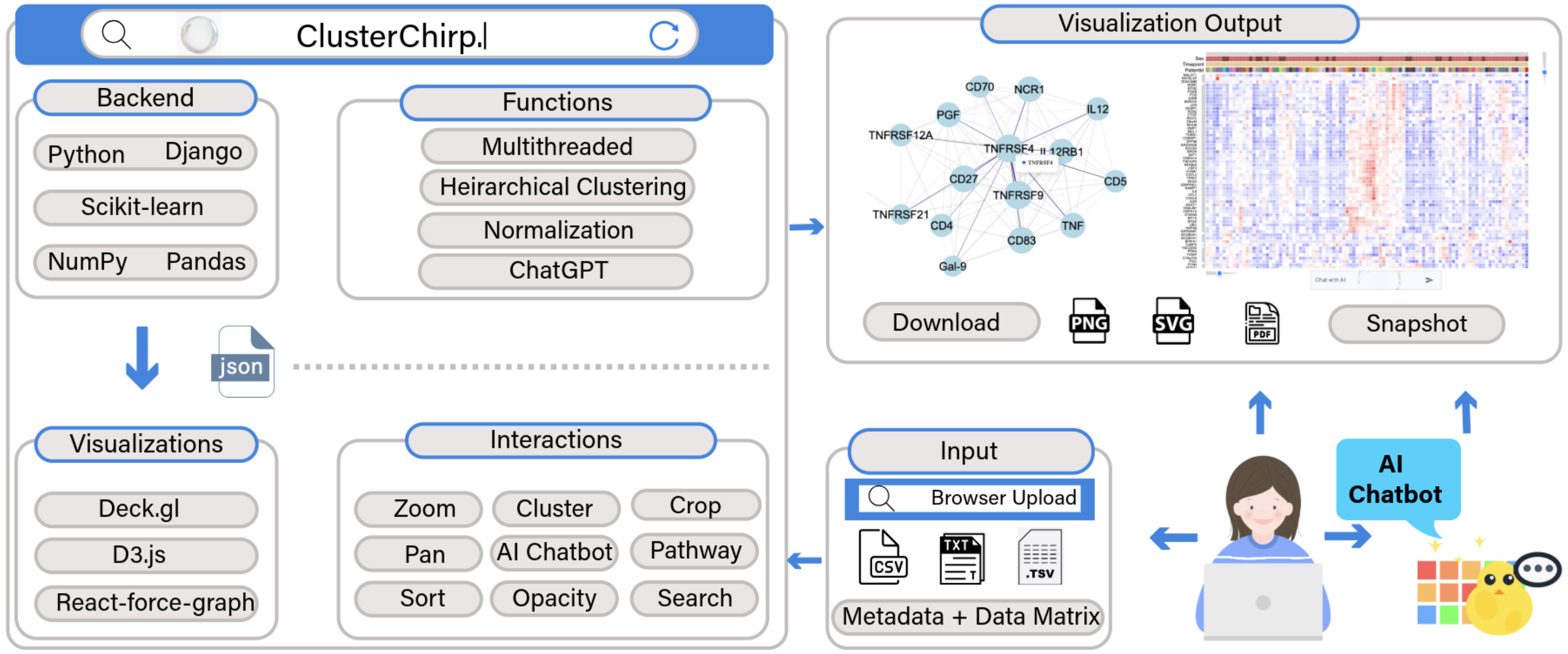

**Figure 1**. Overall architecture of ClusterChirp.

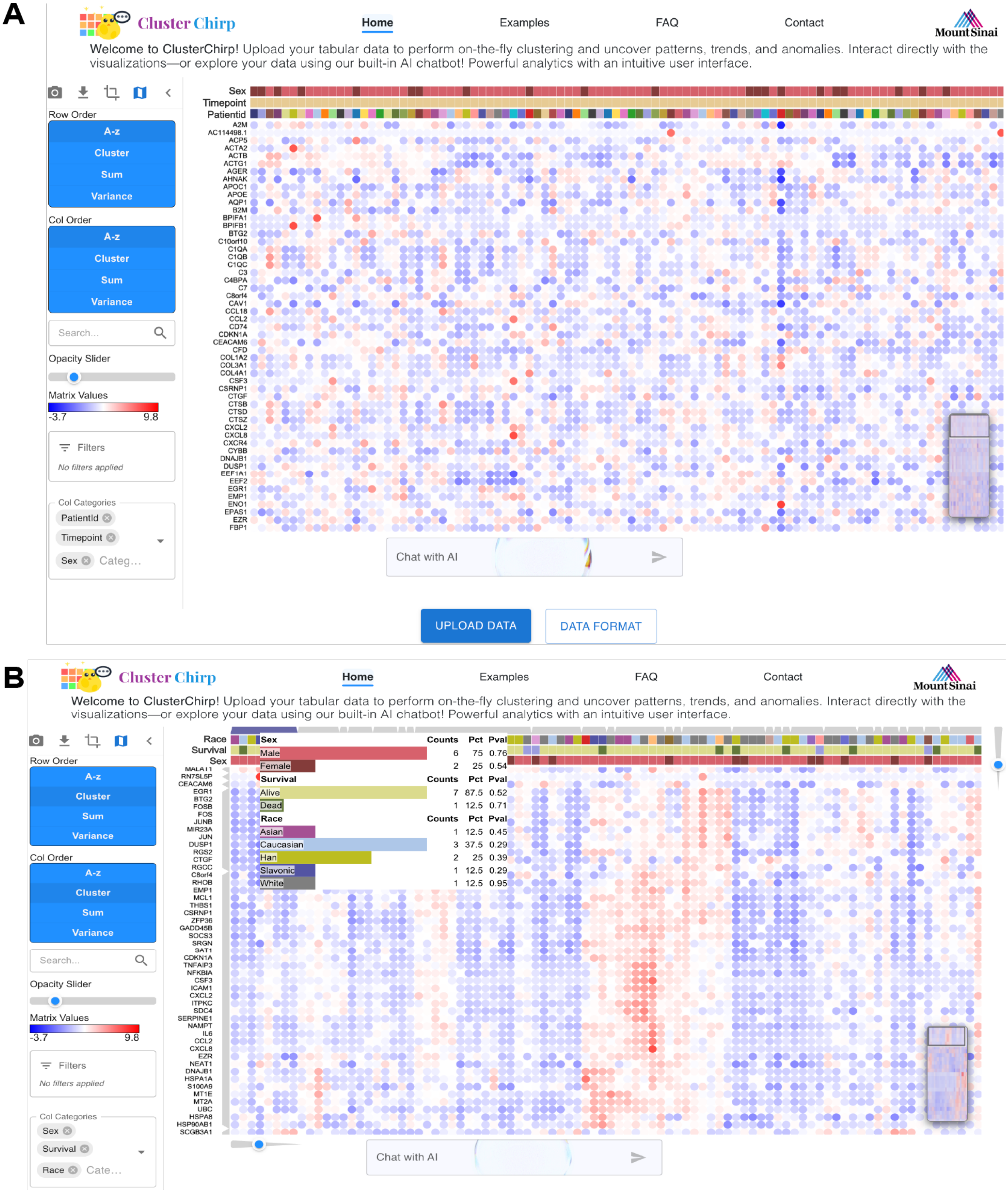

**Figure 2. ClusterChirp web interface. (A) Homepage displaying an example dataset.** The top navigation bar spans tabs for Home, Examples, FAQ (questions and tutorials), and Contact. The left control panel provides options for row and column ordering, search, opacity, value scaling, and filtering. Filters are populated

dynamically as users enter natural language commands in the AI chat box in the bottom; for example, typing "filter top 100 most variant rows" adds a row filter, with column filters generated similarly. Dropdown menus allow selection of up to three row or column annotation categories, automatically detected from the uploaded dataset, and rendered as metadata bars on the heatmap; here, Patient ID, Timepoint, and Sex are selected. Data can be uploaded via the "Upload Data" button, with formatting instructions available under "Data Format". **(B) Heatmap generated after hierarchical clustering**. Row and column dendrograms are displayed alongside the heatmap; interactive sliders enable traversal across clustering depths to visualize clusters at varying resolutions. Hovering over a cluster reveals a metadata summary box with counts, percentages, and p-values for each annotation category; here, Sex, Survival, Race.

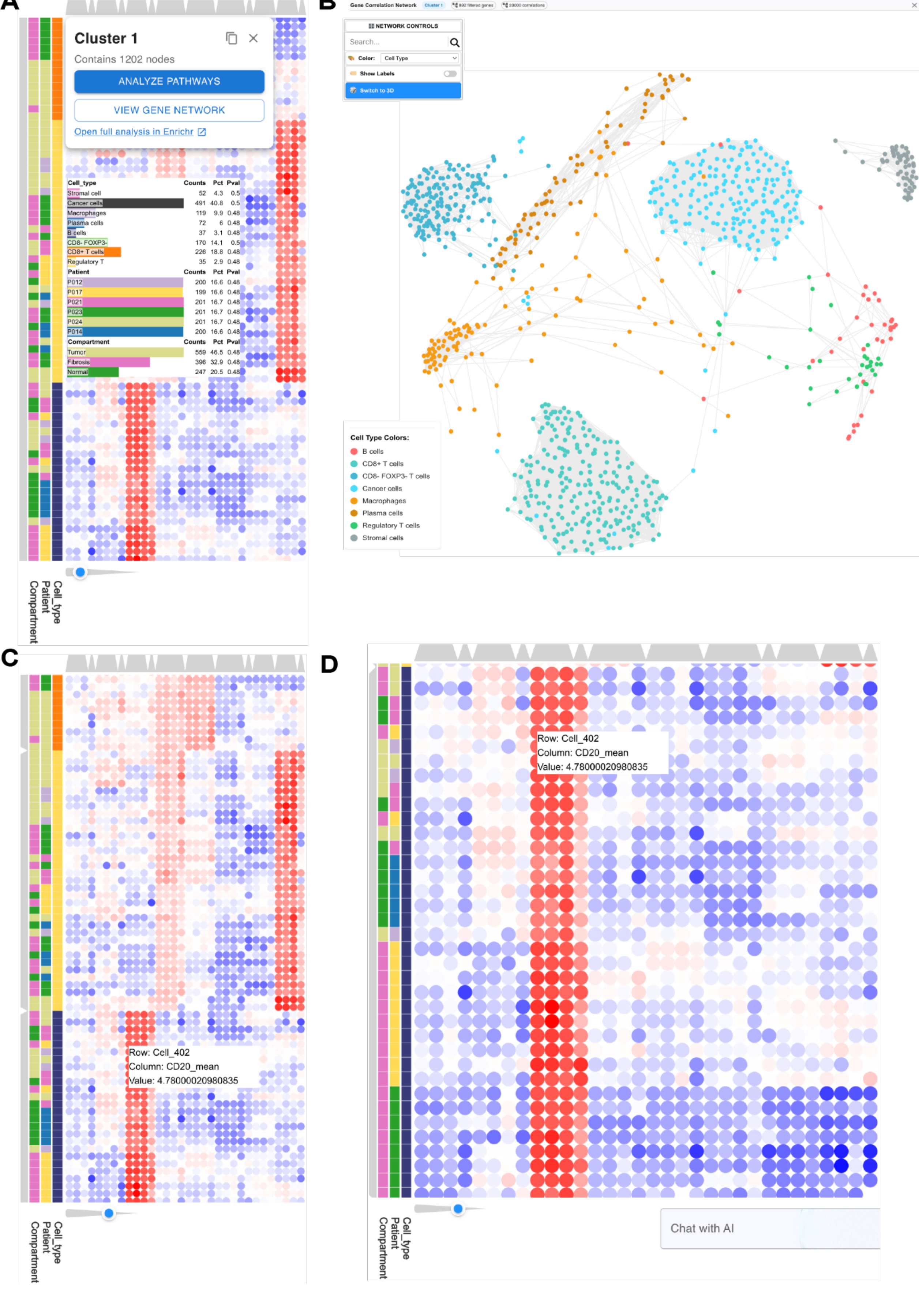

**Figure 3. Spatial proteomics analysis reveals distinct cell type signatures in the tumor microenvironment.** A multiplexed immunohistochemistry (mIHC) dataset of 1,202 single cells from six cancer patients, comprising eight protein markers each quantified at four intensity levels, producing 32 features per cell type across eight annotated cell types. **(A) Hierarchical clustering results with interactive cluster selection.** Hovering over a dendrogram branch displays cluster statistics (cell counts, proportions, p-value) across annotations (Cell Type, Patient, Compartment); clicking opens a dialog with options for pathway enrichment via Enrichr API (34–36), correlation network visualization or direct upload to Enrichr. Red indicates high expression; blue indicates low. **(B) Force-directed network of all 1,202 cells, accessed via the "View Gene Network".** Nodes represent individual cells colored by cell type, with layout determined by marker expression similarity, naturally separating distinct cell populations. **(C) Dendrogram depth slider for multi-resolution cluster exploration.** The slider (bottom left, blue) traverses across 11 hierarchical levels, revealing progressively finer structure; at level 5, distinct cell types resolve into separate clusters. Hovering over any cell displays its row ID, column ID, and exact value. **(D) Subset analysis of B cells using the crop function.** The resulting heatmap shows uniformly high CD20 expression (red columns), confirming B cell identity.

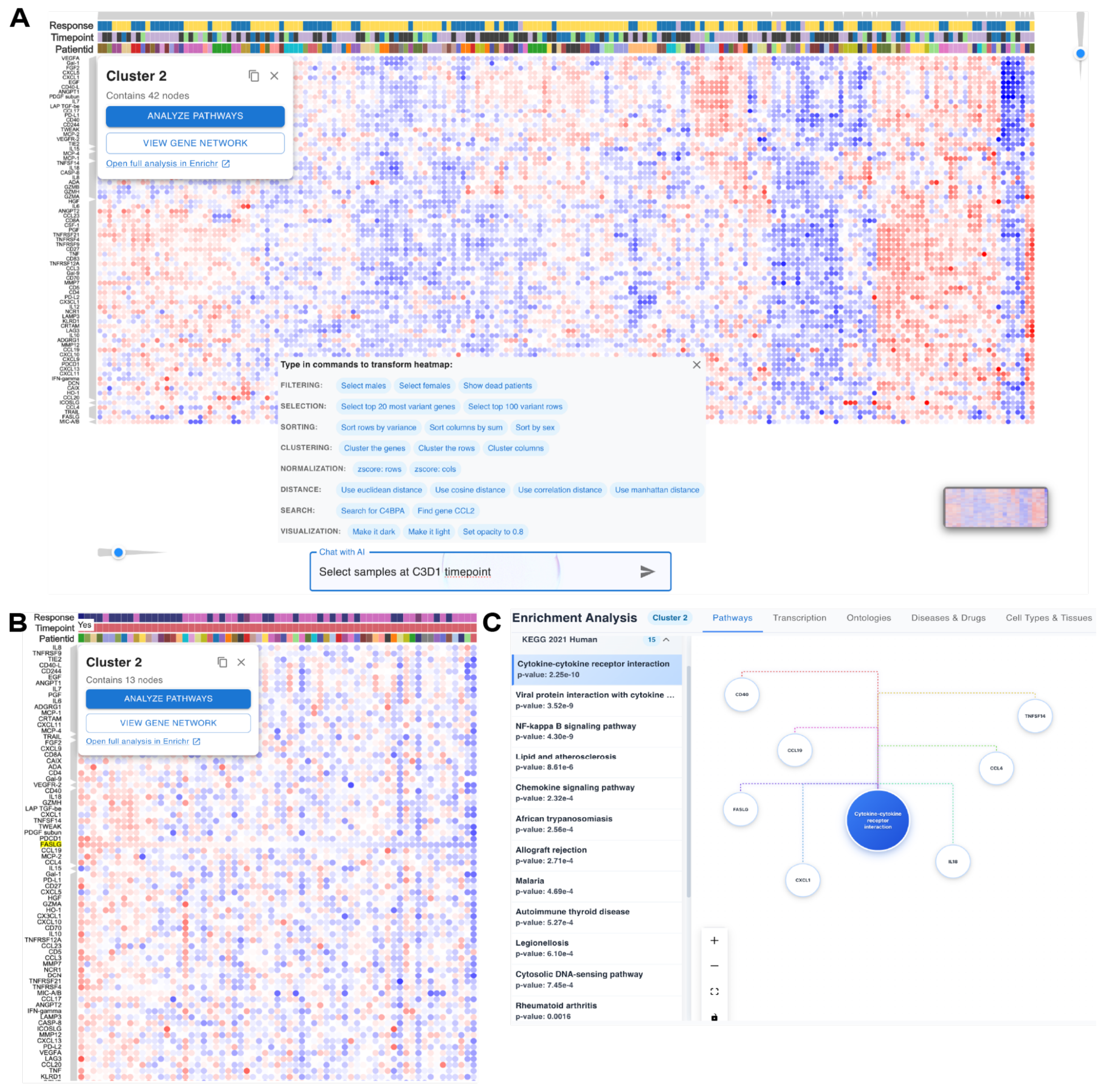

**Figure 4. Natural language-guided analysis of treatment response biomarkers in bladder cancer plasma proteomics.** Data from the GU16-257 bladder cancer immunotherapy trial (42) comprising 77 plasma proteins measured across 196 samples at four treatment cycles. **(A) Hierarchical clustering of the full dataset with cluster selection dialog** (Cluster 2, 42 proteins). The command guide popup lists available natural language operations by category: Filtering (e.g., "Select males"), Selection (e.g., "Select top 100 variant rows"), Sorting (e.g., "Sort rows by variance") and more options, with the AI chat box showing an example command: "Select samples at C3D1 timepoint". Column metadata bars display Response status,

Timepoint, and Patient ID. **(B) Filtered heatmap restricted to C3D1 timepoint samples after AI-assisted filtering**. Rows are sorted by *FASLG* expression (descending). Responders (dark blue, Response bar) are enriched among samples with high *FASLG* expression; Cluster 2 contains 13 proteins in this view. Clicking on a dendrogram branch enables downstream pathway or network analysis or direct export to Enrichr (34–36). **(C)** Pathway enrichment analysis of Cluster 2 using the KEGG 2021 Human database. The top enriched pathway is cytokine-cytokine receptor interaction ($p = 2.25e-10$). The network visualization (right) shows cluster genes associated with this pathway (CD40, CCL19, FASLG, CXCL1, IL1B, CCL4, TNFSF14).

**Table 1. Benchmarking ClusterChirp with other web-based tools.**

| Feature | Morpheus | NG-CHM | Heatmapper2 | ClustVis | Clustergrammer | ClusterChirp |
|---|---|---|---|---|---|---|
| Availability | ✓ | ✓ | ✓ | ✓ | ✓ | ✓ |
| **Size & Performance** | | | | | | |
| Max Matrix Size | No limit (fails) | 5,000 rows | No limit (client RAM) | 2,400 rows | Unknown (server error) | Up to 10M cells |
| Clustering (5K × 2K) | X Running for (12+ min) | 7+ min | ~30 sec | x File size limit | x 502 Server Error | 3 min |
| Clustering (10.9K × 108) | x Error (3.5 min) | x Row limit | ~25 sec | x Row limit | x 502 Server Error | 30 sec |
| **Input & Data** | | | | | | |
| Missing Value Imputation | x | Basic (5) | x | x | x | Advanced (8) + auto |
| **Visualization** | | | | | | |
| Interactive (zoom/pan/hover) | ✓ | ✓ | x Static | x Static | ✓ | ✓ |
| **Interactive Features** | | | | | | |
| Metadata Filtering | ✓ (GUI) | x | x | x | ✓(GUI) | ✓(NLP) |
| Natural Language Interface | x | x | x | x | x | ✓ |
| GPU Acceleration | x | x | x | x | x | ✓ |
| **Downstream Analysis** | | | | | | |
| Correlation Network | x | x | x | x | x | ✓ (2D/3D) |
| Pathway Enrichment | x | x | x | x | ✓(via Enrichr) | ✓ |
| PCA Analysis | x | x | x | ✓ | x | x |
| **Architecture** | | | | | | |
| Parallelized Clustering | x | x | x | x | x | ✓ |
| **Accessibility** | | | | | | |
| No Login Required | ✓ | ✓ | ✓ | ✓ | ✓ | ✓ |
| Actively Maintained | ✓ | ✓ | ✓ | ✓ | ✓ | ✓ |

## Supplementary Tables

### Table S1. Natural language commands that were tested. Total Commands (n=45)

| # | Command Type | Command | Expected Action |
|---|---|---|---|
| **1** | Filtering | show only male samples | Filter Gender = M |
| **2** | Filtering | filter to female samples | Filter Gender = F |
| **3** | Filtering | select responders only | Filter Response = Yes |
| **4** | Filtering | show non-responders | Filter Response = No |
| **5** | Filtering | filter to NE response | Filter Response = NE |
| **6** | Filtering | select samples at C1D1 timepoint | Filter Timepoint = C1D1 |
| **7** | Filtering | show only C3D1 samples | Filter Timepoint = C3D1 |
| **8** | Filtering | filter to C8D1 | Filter Timepoint = C8D1 |
| **9** | Filtering | select C12D1 timepoint | Filter Timepoint = C12D1 |
| **10** | Filtering | show white patients only | Filter Race = white |
| **11** | Filtering | filter to asian samples | Filter Race = asian |
| **12** | Filtering | select black patients | Filter Race = black |
| **13** | Filtering | show hispanic samples | Filter Ethnicity = hispanic |
| **14** | Filtering | filter to non-hispanic | Filter Ethnicity = non_hispanic |
| **15** | Filtering | exclude male samples | Inverse filter Gender ≠ M |
| **16** | Filtering | remove C1D1 timepoint | Inverse filter Timepoint ≠ C1D1 |
| **17** | Filtering | show male responders | Combined: Gender=M AND Response=Yes |
| **18** | Filtering | filter to female samples at C3D1 | Combined: Gender=F AND Timepoint=C3D1 |
| **19** | Filtering | show top 50 variant proteins | Variance-based filter |
| **20** | Filtering | clear all filters | Reset all filters |
| **21** | Clustering | cluster rows | Default row clustering |
| **22** | Clustering | cluster columns | Default column clustering |
| **23** | Clustering | cluster using Pearson correlation | Distance metric = Pearson |
| **24** | Clustering | cluster rows with Euclidean distance | Distance metric = Euclidean |
| **25** | Clustering | cluster using Manhattan distance | Distance metric = Manhattan |
| **26** | Clustering | cluster with cosine similarity | Distance metric = Cosine |
| **27** | Clustering | cluster using average linkage | Linkage = Average |
| **28** | Clustering | cluster with complete linkage | Linkage = Complete |
| **29** | Clustering | cluster rows using single linkage | Linkage = Single |
| **30** | Clustering | cluster genes using Pearson correlation and average linkage | Combined: Pearson + Average |
| **31** | Clustering | cluster samples by Euclidean distance and complete linkage | Combined: Euclidean + Complete |
| **32** | Clustering | cluster columns by Gender | Metadata-based clustering |
| **33** | Clustering | cluster samples by Timepoint | Metadata-based clustering |
| **34** | Clustering | cluster columns by Response | Metadata-based clustering |
| **35** | Sorting | sort rows alphabetically | Alphabetical sort (rows) |
| **36** | Sorting | sort columns alphabetically | Alphabetical sort (columns) |
| **37** | Sorting | sort proteins by variance | Variance-based sort |
| **38** | Sorting | sort samples by sum | Sum-based sort |
| **39** | Sorting | sort rows by variance descending | Descending variance sort |
| **40** | Sorting | sort samples by FASLG expression | Sort by specific protein |
| **41** | Sorting | sort columns by IL8 expression | Sort by specific protein |
| **42** | Sorting | sort samples by Gender | Metadata-based sort |
| **43** | Sorting | sort columns by Timepoint | Metadata-based sort |
| **44** | Sorting | sort samples by Response | Metadata-based sort |
| **45** | Sorting | sort columns by Race | Metadata-based sort |

**Table S2. Performance metrics for the natural language AI interface.**

| Metric Category | Metric | Result |
|---|---|---|
| **Command Success** | Overall Success Rate | 95.66% |
| | Filtering Commands | 96% |
| | Clustering Commands | 98% |
| | Sorting Commands | 93% |
| **Response Time Performance** | Average Command Processing Time | 2 sec |
| | Clustering Operation Time | 30 sec (For 10,000 rows) |
| **Reliability** | Commands Requiring Rephrasing | 4% |
| | API Fallback Rate | 6% |